\documentclass[preprint,showpacs,preprintnumbers,
amsmath,amssymb]{revtex4-1}

\usepackage{graphicx}
\usepackage{dcolumn}
\usepackage{bm,morefloats,color}

\usepackage{amssymb,bm}

\def\Eq#1{\begin{equation} #1 \end{equation}}
\def\Eqr#1{\begin{eqnarray} #1 \end{eqnarray}}
\def\Eqrsubl#1#2{\begin{subequations}\label{#1}\Eqr{#2}\end{subequations}}

\newcommand{\nn}{\nonumber}
\newcommand{\pd}{\partial}

\newcommand{\bea}{\begin{eqnarray}}
\newcommand{\eea}{\end{eqnarray}}

\def\Ysp{{\rm Y}}

\def\X5sp{{\rm X}_5}
\def\Y3sp{{\rm Y}_3}
\def\Z3sp{{\rm Z}_3}

\def\lap{{\triangle}}
\def\e{{\rm e}}

\begin{document}

\title{
Orbifold black holes
}

\author{Muneto Nitta}%
\affiliation{%
Department of Physics, and Research and Education 
Center for Natural Sciences, Keio University, 
Hiyoshi 4-1-1, Yokohama, Kanagawa 223-8521, Japan}%

\author{Kunihito Uzawa}
\affiliation{%
Department of Physics,
School of Science and Technology,
Kwansei Gakuin University, Sanda, Hyogo 669-1337, Japan 
and
\\
Research and Education 
Center for Natural Sciences, Keio University, 
Hiyoshi 4-1-1, Yokohama, Kanagawa 223-8521, Japan
}%

\date{\today}

\begin{abstract}
We construct a regular black hole solution 
on the orbifold ${\mathbb C}^{n}/{\mathbb Z}_{n}$ 
in the ($2n+1$)-dimensional Einstein-Maxwell theory. 
The event horizon is $S^{2n-1}/{\mathbb Z}_{n}$.
\end{abstract}


\maketitle


\section{Introduction}
Black holes give profound conceptual implications for 
the nature of general relativity in both the classical 
and quantum domains. One expects that they will eventually 
play an equally profound role in understanding the 
fundamental nature of particle physics as well as general 
relativity. Accordingly, 
it is important to understand the types of black holes 
which can couple to matter fields. 
If we consider a star which undergoes gravitational 
collapse and form a black hole, we would expect all 
the matter present to be absorbed by the black hole. 
Since the background geometry at sufficiently 
late times should settle down to final state of 
gravitational collapse and  
vacuum except for presence of gauge fields 
associated with the black hole, it is of great 
interest to find all solutions of the Einstein-gauge 
field equations which describe stationary black holes.  
In four space-time dimensions, 
it is known that black holes are {\it unique} 
except for small numbers of the charges. 
The Kerr solution \cite{Kerr:1963ud} 
has angular momentum
in addition to the mass.
Another simplest case is
an electrically charged generalization of  
the Schwarzschild solution in a four-dimensional 
Einstein-Maxwell theory, as given by Reissner-Nordstr\"{o}m 
\cite{Reissner, Weyl, Nordstrom, Jeffery}. 
When the electric charge is proportional to the mass, 
it is called an extremal black hole, 
playing important roles in string theory 
\cite{Townsend:1997ku, Stelle:1998xg, Duff:1999rk}

However, other than four dimensions, 
no such uniqueness theorem is known 
and thus exploring black hole solutions in higher dimensions 
is one of very important problems in general relativity.
In fact, other than black holes, there are also black rings 
\cite{Emparan:2001wn, Elvang:2003yy, Elvang:2004rt, Emparan:2006mm, 
Bena:2007kg} 
for which singularity is not pointlike but ringlike.
Concentrating on black holes,   
one of interesting class of solutions in five dimensions 
is given by 
black holes based on the Eguchi-Hanson space 
\cite{Eguchi:1978xp,Eguchi:1980jx} 
found in 
Refs.~\cite{Ishihara:2006pb,Tatsuoka:2011tx}. 
In this case, the geometry is direct product of 
the time direction ${\mathbb R}$ and 
the Eguchi-Hanson space 
whose center the black hole sits on. 
Multiple black hole solutions were also constructed 
in Ref.~\cite{Ishihara:2006iv}. 
Einstein(-Maxwell) theory with a Gauss-Bonnet term 
also admits similar black hole solutions
\cite{Dehghani:2005zm, Dehghani:2006aa}.

The Eguchi-Hanson space is reduced to 
the orbifold ${\mathbb C}^2/{\mathbb Z}_2$ 
in a certain limit. 
In other words, such an orbifold singularity is 
resolved for the Eguchi-Hanson space. 
This was generalized to the Gibbons-Hawking metric 
\cite{Gibbons:1979zt,Gibbons:1979xm}
on resolved orbifolds ${\mathbb C}^2/{\mathbb Z}_n$ 
  or more generally asymptotically locally Euclidean (ALE) spaces.
In the context of string theory, 
D-branes on orbifolds 
 ${\mathbb C}^2/{\mathbb Z}_n$ 
were studied \cite{Douglas:1996sw}, 
where oribifold singularities are naturally resolved 
in D-barne world-volume theories. 
This was generalized to
 higher dimensional orbifolds 
${\mathbb C}^n/\Gamma$ 
with some discrete groups $\Gamma$ 
\cite{Douglas:1997de}. 
On the other hand, the Eguchi-Hanson space 
can be regarded as 
a Ricci-flat K\"{a}hler manifold
as a cotangent bundle or complex line bundle 
over a sphere $S^2$, 
which is equivalent to ${\mathbb C}{\rm P}^1$.
As a higher dimensional generalization,
the complex line bundle over 
${\mathbb C}{\rm P}^n$ admits a Ricci-flat K\"{a}hler metric 
\cite{Higashijima:2001vk,
Higashijima:2001fp,
Higashijima:2002px} 
and this is obtained by resolving the orbifold singularity in 
the higher dimensional orbifold ${\mathbb C}^{n}/{\mathbb Z}_{n}$. 
This 
implies possible existence of a higher dimensional 
generalization of five dimensional black holes in 
Refs.~\cite{Ishihara:2006pb,Tatsuoka:2011tx}. 
In fact, the authors in 
Refs.~\cite{Tatsuoka:2011tx, Dehghani:2005zm, Dehghani:2006aa} 
considered black hole-like  
solutions based on ${\mathbb C}{\rm P}^n$ but they are singular 
-- a black hole singularity is not surrounded by an event horizon.

In this paper,  
we show that the $D(=2n+1)$-dimensional 
Einstein-Maxwell 
theory admits regular black hole solutions 
on the orbifold ${\mathbb C}^{n}/{\mathbb Z}_{n}$ 
whose orbifold singularity is resolved,  
which is a complex line bundle over ${\mathbb C}{\rm P}^n$.
The black hole singularity is surrounded by an event horizon 
$S^{2n-1}/{\mathbb Z}_n$ 
and thus it is a regular black hole.
This black hole is also electrically charged, 
and the charge is proportional to the mass, 
thereby being an extremal black hole.

In sec.~\ref{EM}, we give an action of the Einstein-Maxwell 
theory. We derive solutions which describe charged black 
holes in this theory. The Einstein-Maxwell theory has a 
2-form field strength which gives rise to charged 
black holes. In sec.~\ref{cpn}, we specialize 
these solutions to the cases of interest for black holes on 
${\mathbb C}^{n}/{\mathbb Z}_{n}$\,. 
We will see that the case of $n=2$ reduces to previously 
found solutions in five dimensions 
\cite{Ishihara:2006pb,Tatsuoka:2011tx}. 
The regularity at the horizon 
of a black hole on the orbifold 
is also discussed. Our solution in this 
paper is produced 
describing a black hole without naked singularity in 
$D(>3)$-dimensional space-time. 
Finally, we conclude with a discussion of our results 
and some implications. 
In appendix A we summarize some general 
results, including an iterative construction of real 
metrics on $\mathbb{C}$P${}^n$, which are needed for
the results in section \ref{cpn}. 

%
\section{$D$-dimensional Einstein-Maxwell theory}
\label{EM}
In this section, we consider Einstein-Maxwell theory in 
$D$ dimensions. We write down the Einstein equations under 
a certain metric ans\"{a}tze, which is a generalization
of those of known static 5-dimensional solutions. 
After we then solve the Einstein equations and present the 
solutions explicitly, we compare the results of $D$-dimensional 
solutions with 5-dimensional one. 

Let us consider a gravitational theory with the metric $g_{MN}$\,, 
and 2-form field strength $F_{(2)}=dA_{(1)}$\,, 
coming from 1-form gauge potential $A_{(1)}$. 
The action for the Einstein-Maxwell theory is written as
\Eq{
S=\frac{1}{2\kappa^2}\int d^Dx\sqrt{-g}\left[R
 -\frac{1}{2\cdot 2!}F_{(2)}^2\right],
\label{EM:action:Eq}
}
where $\kappa^2$ is the $D$-dimensional gravitational constant. 

The field equations are given by 
\Eqrsubl{EM:equations:Eq}{
&&\hspace{-1cm}R_{MN}=\frac{1}{2\cdot 2!} 
\left[2F_{MA} {F_N}^{A}-\frac{1}{D-2}g_{MN} F^2_{(2)}\right],
   \label{EM:Einstein:Eq}\\
&&\hspace{-1cm}d\left[\ast F_{(2)}\right]=0\,,
   \label{EM:gauge:Eq}
}
where $\ast$ is the Hodge dual operator in the
$D$-dimensional spacetime. 

We look for solutions whose spacetime metric has the form
\Eq{
ds^2=-h^{-2}(y)dt^2+h^{\frac{2}{D-3}}(y)u_{ij}(\Ysp)dy^i\,dy^j\,,
\label{EM:metric:Eq}
}
where $u_{ij}(\Ysp)$ is the metric of the $(D-1)$-dimensional 
space Y which depends only on the $(D-1)$-dimensional 
coordinates $y^i$\,, and the function $h$ depends only on $y^i$\,.
We also assume that the gauge field strength $F_{(2)}$ is 
given by
\Eq{
F_{(2)}=\pm\,\sqrt{2}\,\left(1+\frac{1}{D-3}\right)^{1/2}\,
d\left(h^{-1}\right)\wedge dt\,. 
    \label{EM:F:Eq}
}

Under the assumptions given above, we first reduce the gauge field 
equations other than the Einstein equations to a simple set 
of equations. For the 2-form field strength $F_{(2)}=dA_{(1)}$\,, 
it implies that $F_{(2)}$ is a closed form and the Bianchi identity 
is automatically satisfied. 
Also the equation of motion for the gauge field becomes
\Eq{
\lap_\Ysp h=0\,,
   \label{EM:gauge1:Eq}
}
where $\lap_\Ysp$ is the Laplacian with respect to the metric
$u_{ij}(\Ysp)$\,.
In order to complete the system of equations, 
we also have to consider the Einstein equations 
(\ref{EM:Einstein:Eq}). 
Under our ansatz, these equations become
\Eqrsubl{EM:cEin:Eq}{
&&h^{-3-\frac{2}{D-3}}\lap_{\Ysp}h=0\,,
     \label{EM:cEin-tt:Eq}\\
&&R_{ij}(\Ysp)-\frac{1}{D-3}h^{-1}u_{ij}\lap_{\Ysp}h=0\,.
     \label{EM:cEin-ij:Eq}
}
In terms of Eq.~(\ref{EM:gauge:Eq}), 
Eq.~(\ref{EM:cEin:Eq}) are equivalent to \cite{Myers:1986rx}
\Eq{
R_{ij}(\Ysp)=0\,,~~~~~\lap_{\Ysp}h=0\,.
  \label{EM:Ein:Eq}
}

\section{Black holes 
on the orbifold ${\mathbb C}^{n}/{\mathbb Z}_{n}$ 
}
\label{cpn}

Now we apply the general formulation developed 
in the last section to the space which is described as 
a complex line bundle over $\mathbb{C}$P${}^{n-1}$ in order to find 
a generalization of the solution discussed in 
Ref.~\cite{Ishihara:2006pb}. 
The metric $u_{ij}(\Ysp)$ 
on the ($2n$)-dimensional space Y
in 
Eq.~(\ref{EM:metric:Eq}) is thus given by
\Eq{
u_{ij}(\Ysp)dy^idy^j=
dr^2+r^2\left[\left\{d\rho+\sin^2\xi_{n-1}
\left(d\psi_{n-1}+\frac{1}{2(n-1)}\omega_n\right)
\right\}^2+ds^2_{\mathbb{C}{\rm P}^{n-1}}\right],
}
where $r$ is a radial coordinate, 
$\rho$ is a coordinate of $S^1$, 
$\xi_{n-1}$ and $\psi_{n-1}$ are coordinates of 
the $\mathbb{C}{\rm P}^{n-1}$ space 
with the ranges 
$0\le\xi_{n-1}\le \pi/2$\,,~
$0\le \psi_{n-1}\le 2\pi$\,, 
and   
$\omega_{n-1}$
and 
$ds^2_{\mathbb{C}{\rm P}^{n-1}}$ denote 
a one-form and a metric on the  
${\mathbb{C}}{\rm P}^{n-1}$ space, respectively,  
recursively defined as
\cite{Dehghani:2005zm, Dehghani:2006aa, Tatsuoka:2011tx}
\Eqr{
ds^2_{\mathbb{C}{\rm P}^{n-1}}&=&
2n\left[d\xi_{n-1}^2+
\sin^2\xi_{n-1}\,\cos^2\xi_{n-1}
\left\{d\psi_{n-1}+\frac{1}{2(n-1)}
\omega_{n-2}\right\}^2\right.\nn\\
&&\left.+\frac{1}{2(n-1)}\sin^2\xi_n\,
ds^2_{\mathbb{C}{\rm P}^{n-2}}
\right],
} 
and 
\Eqrsubl{cpn:cp1:Eq}{
\omega_{n-2}&=&2(n-1)\sin^2\xi_{n-2}
\left[d\psi_{n-2}+\frac{1}{2(n-2)}\omega_{n-3}
\right],\\
ds^2_{\mathbb{C}{\rm P}^1}&=&4\left(
d\xi_1^2+\sin^2\xi_1\,\cos^2\xi_1\,d\psi_1^2\right)\,,\\
\omega_1&=&4\sin^2\xi_1\,d\psi_1\,.
}
Here, $(r,\rho)$ denotes a complex line,
and $\rho$ together with ${\mathbb C}{\rm P}^n$ 
describe a $2n-1$ dimensional sphere 
$S^{2n-1}/{\mathbb Z}_n = S^{D-2}/{\mathbb Z}_n $, 
which is actually an event horizon.

Let $h$ be a function on Y of the form:
\Eq{
h=h(r)\,. 
}
The remaining non-trivial equation (\ref{EM:Ein:Eq}) reduces to
\Eq{
\lap_\Ysp h=\frac{1}{r^{D-2}}\left(r^{D-2}\,h'\right)'=0\,.
}
Thus, the solution for $h$ under our ansatz
is given by
\Eq{
h(r)=c_0+\frac{c_1}{r^{D-3}}\,,
  \label{cpn:h:Eq}
}
where $c_0$ and $c_1$ are constants. 
The constant of integration $c_0$ 
has been set equal to 1 in the following, 
which is the condition required for having an asymptotically 
flat geometry at $r\rightarrow\infty$, while 
the integration constant $c_1$ in 
Eq.~(\ref{cpn:h:Eq}) 
sets the mass scale of the solution; it has been taken to be 
positive in order to ensure the absence of naked singularities 
at finite $r$\,.

The metric of a spherically symmetric, charged black hole is thus:
\Eq{
ds^2=-\left(
1+\frac{c_1}{r^{D-3}}\right)^{-2}dt^2
+\left(1+\frac{c_1}{r^{D-3}}
\right)^{\frac{2}{D-3}}u_{ij}(\Ysp)dy^i\,dy^j\,.
\label{cpn:metric:Eq}
}
While we compute the physical charge in terms of the 
solution (\ref{cpn:metric:Eq})
\Eq{
Q=\frac{1}{2\kappa^2}\int_{{S}^{D-2}/{\mathbb Z}_n}\ast 
F_{(2)}=\pm\,\sqrt{2(D-2)(D-3)}\,
\frac{c_1\,V_{{S}^{D-2}/{\mathbb Z}_n}}{2\kappa^2}\,,
}
we can also derive the expression of ADM mass and compare it 
with the physical charge 
\Eq{
M=\left(D-2\right)
\frac{
c_1\,V_{{S}^{D-2}/{\mathbb Z}_n}}{\kappa^2}
=\sqrt{\frac{2(D-2)}{D-3}}\,\left|Q\right|\,,
}
where $V_{{S}^{D-2}/{\mathbb Z}_n}$ denotes the volume 
of ${S}^{D-2}/{\mathbb Z}_n$\,. 
Since the mass is proportional to the charge, 
this is an extremal black hole. 

One notes that there are some interesting remarks  
about the solution we have found. 
First, one arises when we note that the function $h$ 
which characterizes the solution diverges at $r=0$. 
For the region of spacetime determined by $r=0$ 
to coincide with the origin of the spherical 
coordinates, and thus with the location of the charged object, 
a necessary condition is that the radius of the 
$(D-2)$-sphere $S^{D-2}$ shrinks to zero size there.
Now the radius of the $S^{D-2}$ in the geometry 
(\ref{cpn:metric:Eq}) is given by $h^{1/(D-3)}r$. 
If we consider qualitatively that the 
function $h$ behaves like $h\sim r^{-(D-3)}$ 
near $r=0$\,, that $h^{1/(D-3)}r\sim$constant\,. 

One would like to know whether the locus $r=0$ is a 
singularity of the 
geometry or not. When this actually can be checked using
the components of the Riemann tensor and the square of 
field strength, 
\Eqrsubl{cpn:2:Eq}{
\lim_{r\rightarrow 0}R_{MNPQ}R^{MNPQ}&=&
-2\,c_1^{2-\frac{4}{D-3}}\,
\left[4D^4-41D^3+154D^2-251D+149\right],\\
\lim_{r\rightarrow 0}F_{MN}F^{MN}&=&
-\frac{2(D-2)}{D-3}c_1^{\frac{2}{D-3}}\,,
}
the invariant does not diverge at $r=0$\,. 
This in turn implies the non-divergence of the square
of the Riemann tensor. 
We may thus conclude that for the charged black hole, 
the curvature does not diverge at the location of the charged 
object itself. This could have been guessed from the 
fact that, for instance, the energy of gauge field is not 
singular at $r=0$\,, and provides a regular source term to 
the Einstein equations. 
However, in order to examine the regularity at 
the horizon $r=0$\,, this conclusion overlooks the study of 
the Riemann curvature measured by a free-falling observer 
with orthonormal bases \cite{Tatsuoka:2011tx}.  
Before drawing conclusions on the physical consequences 
of this regular behaviour, we should consider 
the regularity of the black hole solution in more detail.
We will discuss the regularity of the null hypersurface 
and show the metric components and the Riemann curvature 
components near $r=0$ where is the horizon.  
Then we discuss an analytic extension across the $r=0$ 
surface in our solution.
Since the metric (\ref{cpn:metric:Eq}) is apparently 
singular null surface at $r=0$\,, we will present 
the coordinates so that the surface becomes smooth. 

To see that the $r=0$ surface is smooth, 
let us here introduce the new coordinates 
$(v\,, \tilde{r})$
\Eq{
dv=dt+h^{(D-2)/(D-3)}dr\,,~~~~~
d\tilde{r}=r^{(D-4)}\,dr\,.
} 
The line element turns out to be
\Eq{
ds^2=-h^{-2}dv^2+h^{-\frac{D-4}{D-3}}\,
\tilde{r}^{-(D-4)}\,dv\,d\tilde{r}
+h^{\frac{2}{D-3}}u_{ij}(\Ysp)dy^idy^j\,.
  \label{cpn:metric2:Eq}
}
These coordinates clearly describe smooth
functions of $r$, namely giving an analytic 
extension of the metric (\ref{cpn:metric2:Eq}): 
\Eqrsubl{cpn:c:Eq}{
g_{v\tilde{r}}&=&c_1^{-\frac{D-4}{D-3}}\left[
1-\frac{D-4}{c_1(D-3)}\,r^{D-3}\right]
+O(r^{2(D-3)})
=c_1^{-\frac{D-4}{D-3}}\left(
1-\frac{D-4}{c_1}\,\tilde{r}\right)
+O(\tilde{r}^{2})\,,\\
g_{ij}&=&c_1^{\frac{2}{D-3}}
\left[1+\frac{2}{c_1(D-3)}\,r^{D-3}\right]
+O(r^{2(D-3)})
=c_1^{\frac{2}{D-3}}
\left(1+\frac{2}{c_1}\,\tilde{r}\right)
+O(\tilde{r}^{2})\,.
}
Since these components do not have any 
fractional power of $\tilde{r}$\,, it implies 
the null hypersurface at $r=0$ becomes regular. 

The aim of the rest of this section is to 
discuss the behavior of the geometry near 
$r=0$ in terms of the computation of 
the expansion for an outgoing null
geodesic congruence. Now we compute the curvature for a 
particular metric 
to present an analytic extension for the 
background (\ref{cpn:metric2:Eq}) \cite{Tatsuoka:2011tx}. 
The vielbeins are determined with the 
metric (\ref{cpn:metric2:Eq}): 
\Eqr{
&&e^{(0)}_0 =-1\,,~~~~~e^{(0)}_1 =-h^{\frac{1}{2n}}
\sqrt{h^2-1}\,,~~~~~e^{(1)}_0=-h^{-1}
\sqrt{h^2-1}\,,\nn\\ 
&&e^{(1)}_1=-h^{1+\frac{1}{2n}}\,,~~~~~
e^{(\alpha)}_i=h^{\frac{1}{2n}}\,r\,
\bar{e}^{(\alpha)}_i\,.
   \label{cpn:viel:Eq}
}
Here we choose $e^{(0)\mu}\pd_\mu$ as a geodesic tangent 
vector in the radial direction, 
$\alpha$ runs over the $(D-2)$ space values 
1\,, $\cdots$\,, 
$(D-2)$\,, and the 
vielbeins obey
\Eq{
e^{(0)\mu}\nabla_{\mu}e^{(A)\nu}=0\,,
~~~~~g^{\mu\nu}e^{(A)}_\mu e^{(B)}_\nu=
\eta^{(A)(B)}\,,~~~~u^{ij}
\bar{e}^{(\alpha)}_i\bar{e}^{(\beta)}_j
=\delta^{(\alpha)(\beta)}\,,
}
where $\nabla_\mu$ denotes the covariant derivative 
with respect to the $D$-dimensional metric 
$g_{\mu\nu}$, $\delta^{(\alpha)(\beta)}$ implies 
$(D-2)\times (D-2)$ unit matrix and 
\Eq{
\eta^{(A)(B)}=
\eta_{(A)(B)}={\rm diag}\,
\left(-1, 1, \cdots, 1\right)\,.
} 
The expansion for an outgoing null
geodesic congruence $k^\mu$ is thus 
following \cite{Tatsuoka:2011tx}
\Eq{
\theta_+=h^{\mu\nu}\nabla_\mu k_\nu\,,
    \label{cpn:exp:Eq}
}
where $h^{\mu\nu}$ is defined by
\Eq{
h^{\mu\nu}=g^{\mu\nu}+k^{\mu}\ell^\nu 
+\ell^{\mu}k^\nu\,,
  \label{cpn:exp2:Eq}
}
and the null vector fields $k^\mu$\,, $\ell^\mu$ satisfy
\Eq{
k^\mu k_\mu=\ell^\mu\ell_\mu=0\,,~~~~~
k^\mu\ell_\mu=-1\,.
  \label{cpn:null:Eq}
}
In terms of vielbeins (\ref{cpn:viel:Eq}), 
we take outgoing and ingoing null vector
fields $k^\mu\pd_\mu$ and $\ell^\mu\pd_\mu$ by
\Eqrsubl{cpn:vec:Eq}{
k^\mu\pd_\mu&=&\frac{1}{\sqrt{2}}
\left[e^{(0)\mu}-e^{(1)\mu}\right]\pd_\mu
=\frac{h^2\left(1-h^{-1}\sqrt{h^2-1}
\right)
}{\sqrt{2}}\pd_t
-\frac{h^{-\frac{1}{D-3}}\left(
\sqrt{h^2-1}-h\right)}{\sqrt{2}}\pd_r
\,,\\
\ell^\mu\pd_\mu&=&\frac{1}{\sqrt{2}}
\left[e^{(0)\mu}+e^{(1)\mu}\right]\pd_\mu
=\frac{h^2\left(1+h^{-1}\sqrt{h^2-1}
\right)
}{\sqrt{2}}\pd_t
-\frac{h^{-\frac{1}{D-3}}\left(
\sqrt{h^2-1}+h\right)}{\sqrt{2}}\pd_r
\,.
}
Since the field $h^{\mu\nu}$ vanishes 
$r\ge 0$ with vielbeins 
given in Eq.~(\ref{cpn:viel:Eq}), 
the expansion also becomes zero for $r\ge 0$\,. 
Hence, the hypersurface at 
$r=0$ looks like the apparent horizon for 
observers in the region $r\ge 0$\,. 

We now proceed to the discussion of an analytic extension 
across the horizon $r=0$ briefly, 
and compute the Riemann curvature
measured by a free-falling observer with orthonormal 
bases. The orthonormal frame in the limit $r\rightarrow 0$ are:
\Eqrsubl{cpn:Riemann:Eq}{
R^{(0)(1)(0)(1)}&=&c_1^{-\frac{2}{D-3}}
(D-3)^2+O(r^{2(D-3)})\,,\\
R^{(0)(\alpha)(0)(\beta)}&=&O(r^{2(D-3)})\,,\\
R^{(1)(\alpha)(1)(\beta)}&=&O(r^{2(D-3)})\,,\\
R^{(\alpha)(\beta)(\gamma)(\delta)}&=&
c_1^{\frac{4}{D-3}}
\left(1+\frac{4}{D-3}\frac{r^{D-3}}{c_1}\right)
\delta^{\alpha\beta}
+O(r^{2(D-3)})\,.
}
The other components of the Riemann curvature vanish.  
These components show that  
the solution is not singular at $r=0$ for $D>3$\,. 
Hence we conclude that the horizon $r=0$ is smooth 
and there is ${\rm C}^2$ extension across
the $r=0$ surface. This is 
similar with the result  
that the singularity at the horizon 
disappears for the Reissner-Nordstr\"{o}m metrics. 
\section{Discussions}
  \label{sec:discussions}
We conclude with some comments on the properties of 
the solutions we have discussed in this letter, and 
potential applications. 
We have discussed exact solutions in $D$-dimensional 
Einstein-Maxwell theory since this is the generalization 
of interest for black hole solutions. 
The space-time structure of the geometry has been analyzed, 
and one finds that the locus $r=0$ is a regular horizon. 

The solutions given here display explicitly
the property that the expansion of an outgoing null geodesic
congruence vanishes. It is therefore
likely that there are null hypersurfaces. 
The examination of our solution for black holes on an 
orbifold suggested that these are nonsingular solutions
because the spacetimes look like black holes for an observer 
outside the hypersurface. The essential point is that 
not only the Kretschmann invariant but the 
Riemann curvature in a frame of an observer parallelly 
transported along a free-fall geodesic becomes finite 
for $D>3$\,. Hence, the solution does not have any curvature 
singularity at the null hypersurface. This result leads one 
to conclude that 
the observer along the free-fall geodesic can across the
horizon due to the 
${\rm C}^2$ extension 
traversing there. 

The existence of the electric charge is essential 
in our solutions because they are extremal black holes.
It is an open question whether 
there exist (non-extremal) black hole solutions without an 
electric charge.
It is also interesting problem if our solutions 
can be embedded into supergravity or superstring.
Since our solutions are extremal, it is likely the case that 
they are BPS states preserving a fraction of supersymmetry.

The case of $n=2$ reduces to 
$D=5$ dimensional black holes 
on the resolved orbifold ${\mathbb C}^2/{\mathbb Z}_2$
\cite{Ishihara:2006pb, Tatsuoka:2011tx}.
In this case, multiple black hole solutions were also constructed 
in Ref.~\cite{Ishihara:2006iv}, 
which should correspond to the orbifold ${\mathbb C}^2/
{\mathbb Z}_{k+1}$ for $k$ black holes.
Constructing multi-black holes in general dimensions 
remains as one of future problems.

As a further generalization, 
one could replace 
the ${\mathbb C}{\rm P}^{n-1}$ manifold 
by other homogeneous K\"{a}hler manifolds  
since Ricci-flat metrics on 
homogenous K\"{a}hler manifolds are known
\cite{Higashijima:2001vk,
Higashijima:2001fp,
Higashijima:2002px} 
including 
conifolds 
\cite{Higashijima:2001yn,Higashijima:2001de}.
One would then obtain higher dimensional black holes 
with event horizons of various topologies.

One can clearly modify the derivation in this paper to 
obtain higher-dimensional charged objects in any dimension 
larger than four. 

It is thus an interesting question whether 
or not there exist corresponding 
black $p$-branes in $D$ dimensions 
\cite{Lu:1995cs, Argurio:1997gt}. 
Although we have not dwelt upon  
the aspect of this subject in this paper, 
the reader is referred to Ref.~\cite{Nitta:2020pmp}
for a treatment of some generalization to 
black $p$-brane solutions. 
These includes also $p$-branes on orbifolds, 
and can involve fractional $p$-branes which are 
stucked at
orbifold singularities \cite{Douglas:1996sw, Douglas:1997de, 
Nakajima, Nakajima2, Kimura:2011wh, Eto:2004vy}.

There is an issue to consider the black holes in expanding 
universe. It was pointed out in Refs.~\cite{Binetruy:2007tu, 
Maeda:2009zi}  that we can generalize 
the extremal solutions to time-dependent background, 
while it was noted that the non-extremal  
solutions cannot be extended to a time-dependent solution
\cite{Maeda:2009zi}.

\section*{Acknowledgments}
We thank 
H.~Ishihara,
M.~Kimura, 
and 
S.~Tomizawa 
for useful comments. 
The work of M.N. is supported in part by Grant-in-Aid for Scientific
Research, JSPS KAKENHI Grant Number
JP18H01217. 
The work of K. U. is supported by Grants-in-Aid from the Scientific 
Research Fund of the Japan Society for the Promotion of 
Science, under Contract No. 16K05364. 
This work was supported by the Ministry of Education, Culture,
Sports, Science (MEXT-)Supported Program for the
Strategic Research Foundation at Private Universities
``Topological Science'' (Grant No. S1511006).

\section*{Appendix}
\appendix

\section{Product of $\mathbb{C}$P${}^{n-1}$ space and their U(1) bundles}
\label{sec:CP}
In this section, we discuss 
the Ricci flat space on $\mathbb{C}$P${}^{n-1}$. 
Let us summarize 
the Fubini-Study construction of the 
Einstein-Kahler metric on $\mathbb{C}$P${}^{n-1}$\,.
We introduce the complex coordinates $z^M$
on $\mathbb{C}^{n}$, 
with the flat metric 
\Eq{
ds^2_{\mathbb{C}^{n}}=dz^Md\bar{z}_M\,,
}
and inhomogeneous coordinates 
$\zeta^m=z^m/z^0$\,, in the patch where $z^0
\ne 0$\,. 
Here we split the index $M$ into $M=(0\,, \mu)$\,, 
and $1\le \mu\le n-1$\,. 

If we use \cite{Hoxha:2000jf}
\Eq{
z^0=\e^{i\tau}\left|z^0\right|\,, ~~~~
r=\sqrt{z^\mu\bar{z}_\mu}\,, ~~~~f=1+\zeta^\mu
\bar{\zeta}^{\bar{\mu}}\,,
}
the flat metric on $\mathbb{C}^{n}$ can be 
written by 
\Eq{
ds^2_{\mathbb{C}^{n}}=dr^2+r^2d\Omega_{2n-1}\,,
}
where $d\Omega_{2n-1}$ denotes the metric of the 
$S^{2n-1}$ \cite{Hoxha:2000jf}
\Eqrsubl{cp:sphere:Eq}{
d\Omega_{2n-1}&=&\left(d\tau+A\right)^2
+f^{-1}d\zeta^\mu d\bar{\zeta}^{\bar{\mu}}
-f^{-2} \bar{\zeta}^{\bar{\mu}}\zeta^\nu
d\zeta^\mu d\bar{\zeta}^{\bar{\nu}}\,,\\
A&=&\frac{i}{2}f^{-1}\left(\zeta^\mu
d\bar{\zeta}^{\bar{\mu}}-\bar{\zeta}^{\bar{\mu}}
d\zeta^\mu\right)\,.
}
Since the metric (\ref{cp:sphere:Eq}) denotes the 
unit $S^{2n-1}$ which is described as a U(1) 
bundle over $\mathbb{C}$P${}^{n-1}$\,, 
it contains the Fubini-Study 
metric $ds^2_{\rm FS}$ on $\mathbb{C}$P${}^{n-1}$
\Eq{
ds^2_{\rm FS}=f^{-1}d\zeta^\mu d\bar{\zeta}^{\bar{\mu}}
-f^{-2} \bar{\zeta}^{\bar{\mu}}\zeta^\nu
d\zeta^\mu d\bar{\zeta}^{\bar{\nu}}\,,
}
and is also referred as 
the unit $\mathbb{C}$P${}^{n-1}$ metric. 
The metric (\ref{cp:sphere:Eq}) thus 
corresponds to the hopf fibration of the 
unit $(2n-1)$-sphere. 
For example, if we take $n=3$\,, we get the 
metric of $\mathbb{C}$P${}^2$ space \cite{Gibbons:1978zy,
Gibbons:1979xm}:
\Eq{
ds^2_{\mathbb{C}{\rm P}^2}=\left(1+\rho^2\right)^{-2}d\rho^2
+\frac{\rho^2}{4}\left(1+\rho^2\right)^{-2}\left(
d\psi+\cos\theta d\phi\right)^2
+\frac{\rho^2}{4}\left(1+\rho^2\right)^{-1}\left(
d\theta^2+\sin^2\theta d\phi^2\right)\,,
}
where we introduce coordinates by defining 
Euler angles $\left(\psi\,, \theta\,, \phi\right)$ 
and a radial coordinate $\rho$ \cite{Gibbons:1978zy}
\Eqrsubl{cp:2:Eq}{
\zeta^1&=&\rho\cos\left(\frac{\theta}{2}\right)
\e^{i\left(\psi+\phi\right)/2}\,,\\
\zeta^2&=&\rho\sin\left(\frac{\theta}{2}\right)
\e^{i\left(\psi-\phi\right)/2}\,.
}
There are coordinate singularities at 
$\rho=0$ and $\theta=0$ or $\pi$ if we set 
\Eq{
0\le \theta\le \pi\,,~~~~
0\le \phi\le 2\pi\,,~~~~
0\le \psi\le 4\pi\,,~~~~
0\le \rho\le \infty\,.
}



\end{document}